\documentclass[letterpaper,aps,prl,showpacs,preprint,groupedaddress,floatfix]{revtex4}
\usepackage{amsmath,amssymb}
\usepackage[dvips]{graphics,color,epsfig}

\renewcommand{\Im}{\mathop{\rm Im}}
\renewcommand{\Re}{\mathop{\rm Re}}

\begin{document}
\title{Ultranarrow resonance peaks in the transmission and reflection spectra
    of a photonic crystal cavity with Raman gain}

\author{V.~G.~Arkhipkin}\email{avg@iph.krasn.ru}
 \author{ S.~A.~Myslivets}

\affiliation{%
L.V.~Kirensky Institute of Physics  and \\
Siberian Federal University, Krasnoyarsk, Russia
}%

\date{\today}

\begin{abstract}
The Raman gain of a probe light in a three-state
$\Lambda $-scheme placed into a defect of a one-dimensional
photonic crystal is studied theoretically.  We show that there exists a pump intensity range, where the transmission and reflection spectra of the probe field exhibit \textit{simultaneously} occurring narrow peaks (resonances) whose position is determined by the Raman resonance.  Transmission
and reflection coefficients can be larger than unity at pump intensities of
order tens of $\mu$W/cm$^{2}$. When the pump intensity is outside this
region, the peak in the transmission spectrum turns into a narrow dip.
The nature of narrow resonances is attributed to a drastic dispersion of the nonlinear
refractive index in the vicinity of the Raman transition, which leads to a significant reduction of the group velocity of the probe wave.
\end{abstract}

\maketitle

Micro- and nano-defects in photonic crystals (PC) are capable of
localizing light within a volume smaller than $\lambda ^{3}$ ($\lambda $
being the wavelength) with a high quality-factor of defect modes (see \cite{Bravo.OE, Lalanne.LPhRev}
and references therein). Such structures are often referred to as photonic
crystal cavities or micro- and nano-cavities \cite{Vahala.Nat}.
They find an important application in many different fields such as photonics
\cite{McCall}, nonlinear optics \cite{Bravo.OE}, quantum electrodynamics \cite{Vernooy} and others. These
structures also underlie the design of low-threshold micro- and nanolasers
\cite{Englund} and Raman lasers \cite{Yang, McMillan}. Inserting a resonant medium
(atoms or quantum wells) into a defect results in a significant
modification of spectral properties of the PC \cite{Ivchenko, Khitrova, John}.
Even more intriguing effects can  arise from combining PC properties with the properties  of electromagnetically induced transparency (EIT) \cite{Soljacic.Nat}. It has been shown recently that in a PC with a defect containing a EIT material \cite{Fleisch}, the defect mode Q-factor
for the probe radiation noticeably increases under EIT \cite{Soljacic, Arkhip}, whereas the
width of the transmission spectrum narrows. The increase factor can be of
order $c/v_g\gg 1$ as under EIT it is quite possible that $v_g\ll c$  \cite{Soljacic} ($c$ is the
light velocity in vacuum and $v_{g}$ is the group velocity of a probe wave in
a EIT-medium). A noticeable reduction of the group velocity (slow light)
occurs as well under conventional Raman interaction of a probe
(Raman) radiation with a strong pump (driving field) \cite{Deng, Inouye, Lee, Sharping}, and it occurs
with a smaller loss and over a broader spectral range than under EIT
\cite{Deng}.

\begin{figure}[!h]
\includegraphics[width=0.3\columnwidth]{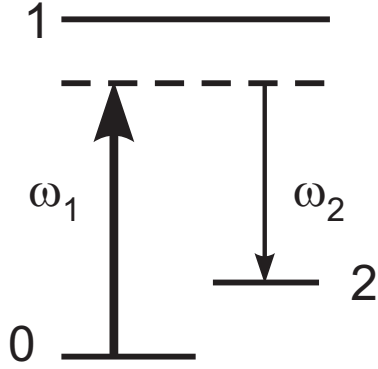}
\caption{\label{fig1}Energy level diagram of a three-level atom in a Raman
gain scheme. States 0 and  2 are the ground and metastable states respectively.}
\end{figure}

In this Letter we suggest a new approach to reduce the
width of the transmission (reflection) spectrum of a PC. Our technique is
based on the effect of Raman gain \cite{Klyshko, Akhmanov} of a probe wave in a defect layer
containing a three-level medium (Fig.~\ref{fig1}). A probe (Raman) wave with frequency $\omega
_{2}$ undergoes amplification when interacting with a coherent pump (driving)  wave with frequency $\omega
_{1}$ as the difference between the two frequencies
comes close to the Raman transition frequency $\omega _{20 }=\omega
_{1}-\omega _{2}$. The pump intensity is chosen so
as to ensure enhancement of the probe wave without however exceeding the
stimulated Raman scattering (SRS) threshold. Here,
unlike spontaneous Raman scattering, phasing of
atomic oscillations occurs throughout the entire volume occupied by light
waves, just as it happens under SRS, but without uncontrollable
 instabilities and with the spectral resolution
being determined by the spectral width of applied
laser radiation. This scheme is weakly  sensitive to pump field intensity
fluctuations \cite{Akhmanov}. We note that Raman gain is being extensively used in high
resolution Raman spectroscopy  in
gases and liquids \cite{Akhmanov} as well as in designing high-efficiency
continous-wave Raman amplifiers and lasers \cite{Repasky, Kumar}.

Consider a one-dimensional photonic crystal having a
(HL)$^{M}$HDH(LH)$^{M }$-type structure. Here $H$ and $L$ refer to different
dielectric layers with a high and low refractive index, $n_{H }$ and
$n_{L}$, and thicknesses $t_{H }$ and $t_{L}$, respectively; D is the defect
layer with a $t_{D}$ thickness and the refractive index $n_{D}$; M is the
number of periods. The defect layer is filled with three-level atoms. Figure~\ref{fig1} shows the energy level diagram and relevant laser couplings for the present study. The concentration of atoms is such that we
can  assume that there is no interaction between atoms. Parameters of
the PC are chosen to allow excitation of just one defect mode
whose spectral width is much larger than that of the allowed transitions and
of the Raman transition $|0\rangle - |2\rangle$ (in contrast to \cite{Yang, McMillan} and references
therein).

Two monochromatic plane waves (the pump and probe) with $\omega_{1,2}$ are normally incident on the PC and
propagate along $z$-axis ($z=0$ at the first layer boundary), which is
perpendicular to the PC layers.
The frequency difference $\omega _{1}-
\omega _{2}$ is close to the Raman transition frequency $\omega _{20}$.
The pump field $E_{1}$ interacts with the $|0\rangle - |1\rangle$ transition and the probe
field $E_{2}$ with the adjacent transition $|1\rangle - |2\rangle$. The
$|0\rangle - |2\rangle$ transition is dipole-forbidden. Only the lower ground
state $|0\rangle$ is initially populated. Both waves are assumed to fall
within the transmission band of the photonic crystal, i.e. the transition
frequency $\omega _{20}$ is less than the defect mode width. For
simplicity, a unity refractive index is assumed for the medium containing
the photonic crystal.

The complex refractive index of the defect layer $n_{D} = n_{2}$ for a probe
field in the presence of a pump wave is given by:
\begin{equation*}\label{eq1}
n_2=n_2'+in_2''=1+2\pi N(\chi_2^{(1)}+ \chi_R|E_1|^2),
\end{equation*}
where $\chi _{2}^{(1)}$ is the linear nonresonant susceptibility for the probe
field;  $E_{1}$ is the complex amplitude of the pump wave; $N$ is the concentration of atoms; $\chi _{R }$ is the Raman susceptibility~\cite{Yariv}
\begin{equation*}\label{eq2}
\chi_R =\frac{1}{4\hbar^3}\frac{d_{21}^2d_{10}^2}{(\omega_{10}-\omega_1)^2
[\omega_{20}-(\omega_1-\omega_2)+i\gamma_{20}]}.
\end{equation*}
Here  $\omega _{10}$
and $\omega _{20}$ are the frequencies of atomic transitions;
$\gamma _{20}$ is the $|0\rangle - |2\rangle$ Raman transition halfwidth;
$d_{ij}$ is the matrix dipole moment of the transition; $\hbar$ is the Plank
constant.

The formula for $\chi _{R}$ was obtained in the third-order of
perturbation theory under the following  conditions: $\Omega_1=|\omega_{10}-\omega_1|\gg |G_{1}|, |G_{2}|, \gamma_{10}$ and $|G_{1}|\gg |G_{2}|$,  $2G_{1}$, and $2G_{2}$ are Rabi frequencies of
the pump and the probe wave, respectively, and $\gamma _{10}$ is the
halfwidth of $|0\rangle - |1\rangle$ transition. Population of the lower state
$|0\rangle$ can be considered unaffected under these conditions.
For simplicity, we neglect the Doppler broadenings  because the one-photon detuning $\Omega_1$ is sufficiently large and the residual Doppler broadening of Raman transition is small.

Note that in the given approximation, $|\Im\chi_R|\gg|\Im\chi_2^{(1)}|$, and the only effect of
$\Re\chi_2^{(1)}$ is to shift the resonant frequency of the defect mode. Therefore the contribution
of $\chi _{2}^{(1)}$ into the refractive index $n_{2}$ will be neglected
in our further consideration. It is essential that the imaginary part of
$\chi _{R}$ is negative in the vicinity of the Raman resonance,
which implies the probe wave enhancement due to energy transfer from the pump to the
probe field. The real part of the refractive index $n'_{2} = \Re{n_{2}}$  has normal dispersion ($dn'_{2}/d\omega_{2}>0$) \cite{Boyd} in this
region, therefore the group velocity of the probe wave can be smaller
than the light velocity in vacuum \cite{Deng}.

In a steady-state approximation, a field in an arbitrary $j-$th layer (\textit{j = H,L,D}) can be treated as
a superposition of counter-propagating waves
\begin{equation*}\label{eq3}
E_j=A_j\exp[ik_j(z-z_j)]+B_j\exp[-ik_j(z-z_j)],
\end{equation*}
where $A_{j}$ and $B_{j}$ are amplitudes of the forward (incident) and
backward (reflected) waves; $k_j=n_j\omega_{1,2}/c$; $n_{j}$ is the
refractive index of  a $j$-th layer. Note that the refractive index of a
defect layer for the probe wave $n'_{2}$ depends
on the spatial coordinate $z$ since distribution of
fields in a defect is non-uniform due to the effect of localization.

Amplitudes $A_{j}$ and $B_{j}$ for each layer were found from wave equations
by means of recurrent relations \cite{Arkhip, Balakin} using the continuity
of tangential components of the electric and magnetic fields
 at the interface of adjacent layers. The transmission and reflection spectra  were determined as
\begin{equation*} \label{eq4}
T(\omega)=|A_{2}(L)| ^2/|A_{02}|^2,
\quad
R(\omega)=|{B_{2}(0)|^2/|A_{02}}|^2,
\end{equation*}
where $A_{02}$ and $A_{2}(L)$ are the input ($z = 0$) and output ($z=L$ is the
photonic crystal length) amplitudes of the probe wave, respectively, and
$B_{2}(0)$ is the amplitude of the probe wave reflected from the input face
of the photonic crystal.

For the purpose of numerical simulation, we used sodium atomic parameters as
a Raman medium. Wavelengths of the probe  and the pump fields were
chosen to be close to $D_1$-line and $\omega_{20}$ to be 1.8 GHz. The photonic
crystal had the following parameters: $M=10$, $n_{H}d_{H} = n_{L}d_{L}=\lambda_{2}/4$,
$d_{D}n_{D}=\lambda_{2}/2$, $n_{H} = 2.35$, $n_{L} = 1.45$. The probe wavelength
was chosen so that its frequency under Raman resonance $\omega
_{1}-\omega _{2}=\omega _{20}$ would match the defect mode
resonance frequency, the pump detuning being $\Omega _{1} = 30\gamma_{10}$, $\gamma_{10}=2\pi\cdot10^8$~s$^{-1}$, $\gamma_{20}/\gamma_{10}=0.1$, $N\simeq10^{12}$~cm$^{-3}$.
For the chosen parameters, calculation of field
distribution in the empty defect layer yields a virtually complete spatial overlapping
of the pump and the probe field.  Intensities of both fields in the defect layer
appear to be $10^{5 }$ times as strong as the input ones. Since we assume
that in a defect layer $|G_1|\gg |G_2|$,
simulation of the transmission and reflection coefficients for the probe
field was performed in the undepleted-pump approximation.

\begin{figure}
\includegraphics[width=0.49\columnwidth]{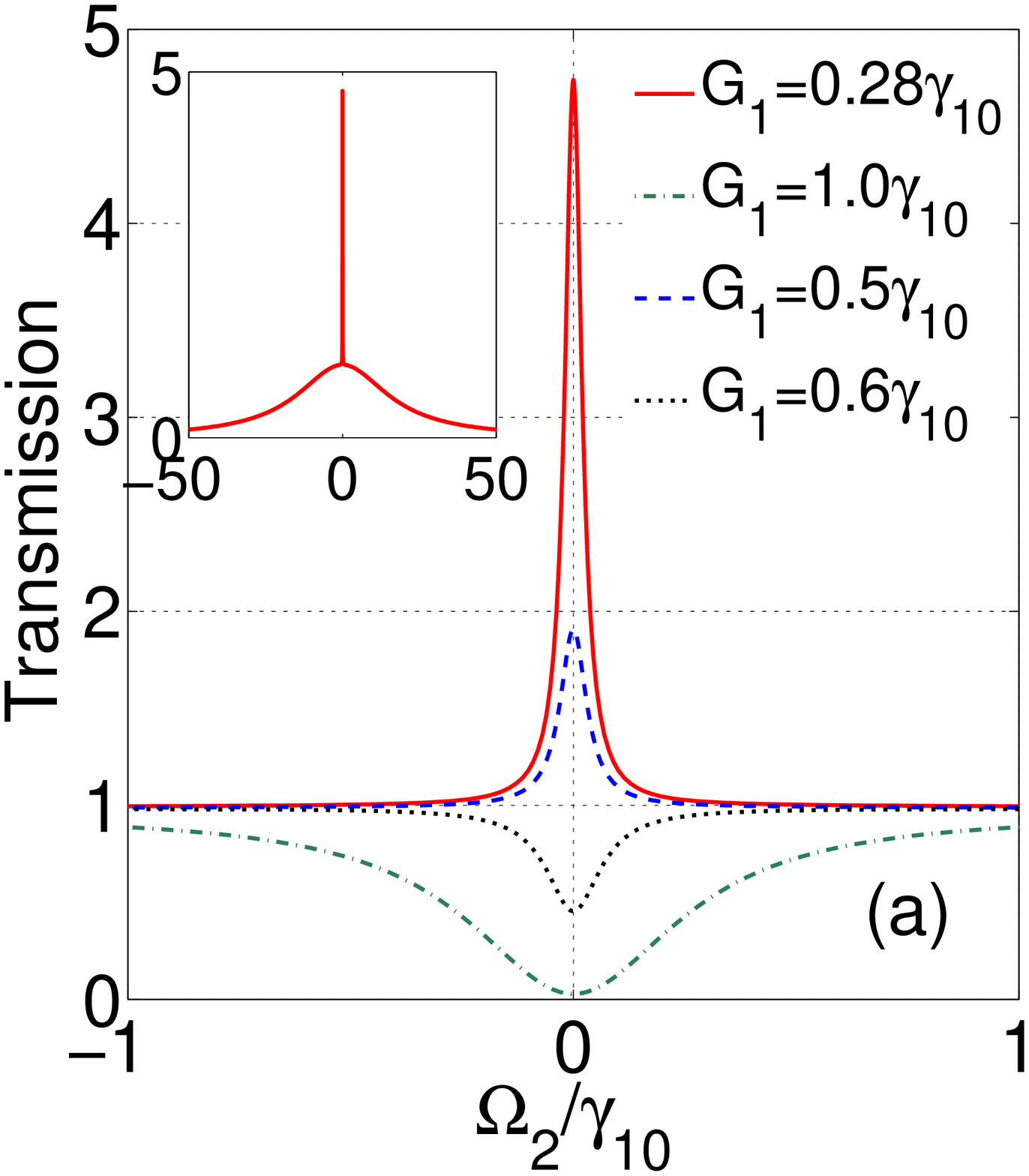}
\includegraphics[width=0.49\columnwidth]{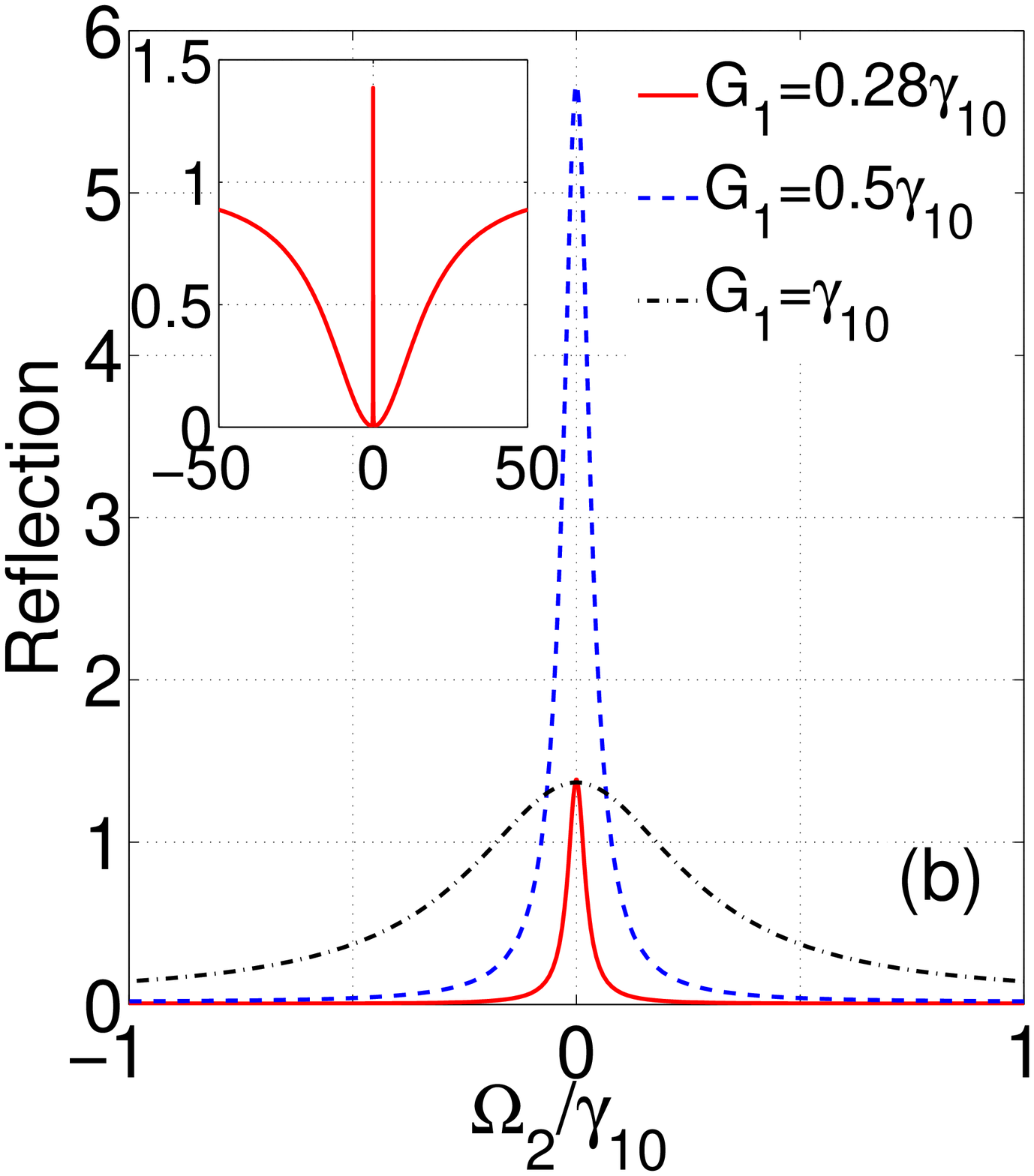}
\caption{\label{fig2}Transmission  $T$ (a) and reflection $R$ (b) spectra   of the
photonic crystal cavity for the probe light vs the
probe field detuning $\Omega _{2} = (\omega _{12 }-\omega _{2})$ scaled to $\gamma _{10}$ for various values of a Rabi frequency of
the pump field $G_{1}$.}
\end{figure}

\begin{figure}
\includegraphics[width=0.49\columnwidth]{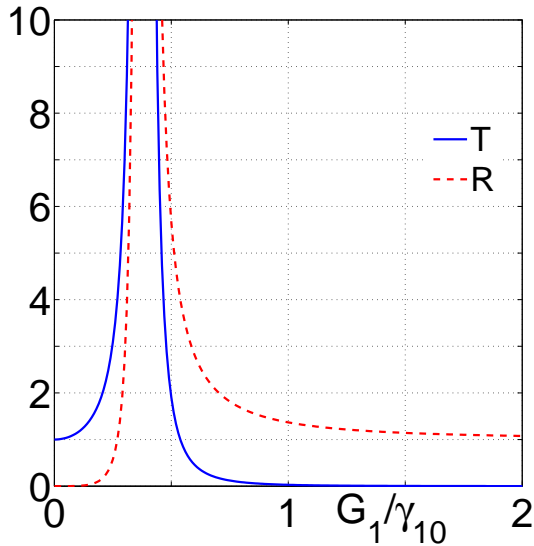}
\caption{\label{fig3}The transmittance and reflectance maxima vs the Rabi frequency of the driving field $G_1$. }
\end{figure}

Figure~\ref{fig2}a shows typical PC transmission and reflection spectra
calculated for the probe wave at different intensities of the driving field. Narrow structures (a peak or a dip) due to the Raman resonance can be observed in the center on the background
of a broad transmission band (Fig.~\ref{fig2}a). The transmittance can be larger than unity. A
narrow peak is also observed in the center of the dip in the reflection
spectrum (Fig.~\ref{fig2}b) and again the reflectance can be larger than unity. So
in a PC with a defect producing the Raman gain, narrow peaks are observed
\emph{simultaneously} in both the transmission and the reflection spectrum. In Fig.~\ref{fig3} transmittance and reflectance maxima are  plotted  as a function of the Rabi frequency of the driving field.  It is seen that the amplitude of the transmission and reflection peaks enhances with the growing pump intensity until the Rabi frequency reaches a threshold, whose value depends on the system parameters. Once this
frequency goes beyond the threshold, the amplitude of  the transmission peak decreases and  the narrow  peak is replaced by a dip while the reflectance tends to unity.

A qualitative interpretation of the features of the transmission
and reflection spectra of a probe field becomes possible if we look at the
problem in terms of a Fabri-Perot cavity (FPC) \cite{Lalanne.LPhRev} with the
length $d$ equal to the thickness of the defect layer $t_{D}$, which is filled
with a Raman medium. The FPC transmittance  for a probe wave $T = I_{2}/I_{20}$
($I_{20}$ is the light intensity as it enters FPC and $I_{2}$ is the
transmitted light intensity) can be found from the
formula
\begin{equation*}\label{eq5}
T=\frac{T_M^2 e^{\alpha d}}{(1-R_M e^{\alpha d})^2 +4R_M e^{\alpha d}\sin^2(\Phi/2)},
\end{equation*}
where $T_{M}$ and $R_{M}$ are the transmission and reflection coefficients of
mirrors; $\alpha = -(4\pi/\lambda )n''_{eff} > 0$ is the
Raman gain factor of the probe wave; $\Phi  = (4\pi /\lambda )n^{'}_{eff}d$ is the phase shift
 after two passes through the cavity; $n''_{eff}=2\pi N\chi_R''F|E_1|^2$ and $n'_{eff}=1+2\pi N\chi_R'F|E_1|^2$ are
effective imaginary and real parts of the refractive index $n_{2}$; $E_{1}$ is
the pump field amplitude in FPC; $F=\int_0^d\sin(k_2z)\sin^2(k_1z)dz/\int_0^d\sin^2(k_2z)dz$ is the spatial overlapping integral of the
pumping wave and the probe field \cite{Repasky1}.

The requirement that $\Phi=2\pi m$ $(m=1,2,\ldots)$ determines a resonance
frequency of the cavity, for which maximum transmission is observed
\begin{equation}\label{eq6}
T_{max}=\frac{T_M^2 e^{\alpha d}}{(1-R_M e^{\alpha d})^2}
\end{equation}
For $\alpha d\ll1$, formula (\ref{eq6}) can be rewritten as
\begin{equation}\label{eq7}
T_{max}\simeq\frac{T_M^2}{(T_M-\alpha d R_M)^2}.
\end{equation}

From formula (\ref{eq7}) it follows that at $\alpha d R_M<T_M$ the transmission
coefficient grows with the pump intensity ($(T_M-\alpha d R_M)\to 0$)  and can
become $T_{max}\gg 1$. In an opposite situation, when $\alpha d R_M>T_M$, the transmittance $T_{max}$ decreases  as the pump field grows (Fig.~\ref{fig3}), and when  $\alpha d R_M > 2T_M$ a
dip is formed in the transmission curve. A similar approach can be applied
to analyze the reflection coefficient.

The width (at half-maximum) of the narrow  transmission peak is given by
the expression:
\begin{equation}\label{eq8}
\delta\omega=\frac{\Delta\omega}{1+\eta}, \qquad
\Delta\omega=\frac{c}{d}\frac{|1-R_M e^{\alpha d}|}{e^{\alpha d/2}\sqrt{R_M}}\simeq\\
\frac{c}{d}\frac{|T_M-\alpha d R_M|}{\sqrt{R_M}},
\end{equation}
where $\eta=2\pi NF|E_1|^2\omega_0 \partial{\chi'_R}/\partial(\omega_2)=K_{12}|G_1|^2/\Omega_1^2\gamma_{20}^2$; $\omega _{0}$ is the resonance frequency of the empty
cavity; $K_{12}=2\pi NF\omega_0|d_{12}|^2/\hbar$. A frequency derivative is
taken at $\omega_2=\omega_0$. For $\alpha  = 0$, the formula for $\Delta \omega $
turns into a familiar expression for a transmission bandwidth of the empty
FPC
\cite{Yariv1}. Note that when $F = 1$, the value of
$1+\eta$ equals the group velocity index $N_{g}=c/v_{g}$ for a probe wave under
Raman interaction \cite{Deng}, which can be much larger than unity if dispersion is
high (at small width of the Raman transition).


It is seen from (\ref{eq8}) that the width of the transmission peak for
$\eta\gg1$ is a factor of $\eta$ narrower than that of the empty cavity, i.e. there
appears a narrow transmission peak. The value of $\eta $ is determined by Raman
susceptibility dispersion and depends on the pump intensity. Analysis
reveals that the resonance width can be less than the width of the Raman
transition $\gamma _{20}$.
The intensity required for
these effects to be observed depends on a number of factors (one-photon pump
frequency detuning, Raman resonance width, quality-factor of defect modes) and can be
anything  from several to hundreds $\mu $W/cm$^{2}$ and less.

To summarize the above, we have studied theoretically light propagation
through a photonic crystal with a defect, filled  with a Raman gain medium  and shown that
narrow peaks can arise simultaneously in the transmission and reflection spectra of the
probe radiation. The position of resonant peaks is determined by the
Raman resonance. Transmission and reflection coefficients can be larger than unity. The
nature of narrow resonances is attributed to dispersion of the nonlinear
refractive index near a Raman transition. Narrow band lasers are required to
be able to observe the described effects. We believe that the predicted effects can be also
observed in cold atoms, including single atoms, placed
into a PC defect, similar to EIT in a cavity
\cite{Shimizu, Zhang}. A combination of the Raman gain effect with the advantages of photonic crystal cavities can be useful for various applications. For example this could help to further reduce the group velocity and obtain longer pulse delays thereby facilitating the designing of Raman lasers of a new generation  and atomic
clock.

This work was supported in part by  the Presidium of the RAS (project no. 27.1),  by the Department of Physical Sciences, RAS (project no. 9.1), and also by  Federal Task Programm (g/c 02.740.11.0220).


\begin{thebibliography}{30}
\expandafter\ifx\csname natexlab\endcsname\relax\def\natexlab#1{#1}\fi
\expandafter\ifx\csname bibnamefont\endcsname\relax
  \def\bibnamefont#1{#1}\fi
\expandafter\ifx\csname bibfnamefont\endcsname\relax
  \def\bibfnamefont#1{#1}\fi
\expandafter\ifx\csname citenamefont\endcsname\relax
  \def\citenamefont#1{#1}\fi
\expandafter\ifx\csname url\endcsname\relax
  \def\url#1{\texttt{#1}}\fi
\expandafter\ifx\csname urlprefix\endcsname\relax\def\urlprefix{URL }\fi
\providecommand{\bibinfo}[2]{#2}
\providecommand{\eprint}[2][]{\url{#2}}

\bibitem[{\citenamefont{Bravo-Abad et~al.}(2007)\citenamefont{Bravo-Abad,
  Rodriguez, Bermel, Johnson, Joannopoulos, and Soljacic}}]{Bravo.OE}
\bibinfo{author}{\bibfnamefont{J.}~\bibnamefont{Bravo-Abad}},
  \bibinfo{author}{\bibfnamefont{A.}~\bibnamefont{Rodriguez}},
  \bibinfo{author}{\bibfnamefont{P.}~\bibnamefont{Bermel}},
  \bibinfo{author}{\bibfnamefont{S.~G.} \bibnamefont{Johnson}},
  \bibinfo{author}{\bibfnamefont{J.~D.} \bibnamefont{Joannopoulos}},
  \bibnamefont{and} \bibinfo{author}{\bibfnamefont{M.}~\bibnamefont{Soljacic}},
  \bibinfo{journal}{Opt. Express} \textbf{\bibinfo{volume}{15}},
  \bibinfo{pages}{16161} (\bibinfo{year}{2007}).

\bibitem[{\citenamefont{Lalanne et~al.}(2008)\citenamefont{Lalanne, Sauvan, and
  Hugonin}}]{Lalanne.LPhRev}
\bibinfo{author}{\bibfnamefont{P.}~\bibnamefont{Lalanne}},
  \bibinfo{author}{\bibfnamefont{C.}~\bibnamefont{Sauvan}}, \bibnamefont{and}
  \bibinfo{author}{\bibfnamefont{J.}~\bibnamefont{Hugonin}},
  \bibinfo{journal}{Laser and Photonics Review} \textbf{\bibinfo{volume}{2}},
  \bibinfo{pages}{514} (\bibinfo{year}{2008}).

\bibitem[{\citenamefont{Vahala}(2003)}]{Vahala.Nat}
\bibinfo{author}{\bibfnamefont{K.~J.} \bibnamefont{Vahala}},
  \bibinfo{journal}{Nature} \textbf{\bibinfo{volume}{424}},
  \bibinfo{pages}{839} (\bibinfo{year}{2003}).

\bibitem[{\citenamefont{McCall et~al.}(1992)\citenamefont{McCall, Levi,
  Slusher, Pearton, and Logan}}]{McCall}
\bibinfo{author}{\bibfnamefont{S.~L.} \bibnamefont{McCall}},
  \bibinfo{author}{\bibfnamefont{A.~F.~J.} \bibnamefont{Levi}},
  \bibinfo{author}{\bibfnamefont{R.~E.} \bibnamefont{Slusher}},
  \bibinfo{author}{\bibfnamefont{S.~J.} \bibnamefont{Pearton}},
  \bibnamefont{and} \bibinfo{author}{\bibfnamefont{R.~A.} \bibnamefont{Logan}},
  \bibinfo{journal}{Applied Physics Letters} \textbf{\bibinfo{volume}{60}},
  \bibinfo{pages}{289} (\bibinfo{year}{1992}).

\bibitem[{\citenamefont{Vernooy et~al.}(1998)\citenamefont{Vernooy, Furusawa,
  Georgiades, Ilchenko, and Kimble}}]{Vernooy}
\bibinfo{author}{\bibfnamefont{D.~W.} \bibnamefont{Vernooy}},
  \bibinfo{author}{\bibfnamefont{A.}~\bibnamefont{Furusawa}},
  \bibinfo{author}{\bibfnamefont{N.~P.} \bibnamefont{Georgiades}},
  \bibinfo{author}{\bibfnamefont{V.~S.} \bibnamefont{Ilchenko}},
  \bibnamefont{and} \bibinfo{author}{\bibfnamefont{H.~J.}
  \bibnamefont{Kimble}}, \bibinfo{journal}{Phys. Rev. A}
  \textbf{\bibinfo{volume}{57}}, \bibinfo{pages}{R2293} (\bibinfo{year}{1998}).

\bibitem[{\citenamefont{Englund et~al.}(2008)\citenamefont{Englund, Altug,
  Ellis, and Vuckovi\'c}}]{Englund}
\bibinfo{author}{\bibfnamefont{D.}~\bibnamefont{Englund}},
  \bibinfo{author}{\bibfnamefont{H.}~\bibnamefont{Altug}},
  \bibinfo{author}{\bibfnamefont{B.}~\bibnamefont{Ellis}}, \bibnamefont{and}
  \bibinfo{author}{\bibfnamefont{J.}~\bibnamefont{Vuckovi\'c}},
  \bibinfo{journal}{Laser and Photonics Review} \textbf{\bibinfo{volume}{2}},
  \bibinfo{pages}{264} (\bibinfo{year}{2008}).

\bibitem[{\citenamefont{Yang and Wong}(2007)}]{Yang}
\bibinfo{author}{\bibfnamefont{X.}~\bibnamefont{Yang}} \bibnamefont{and}
  \bibinfo{author}{\bibfnamefont{C.~W.} \bibnamefont{Wong}},
  \bibinfo{journal}{Opt. Express} \textbf{\bibinfo{volume}{15}},
  \bibinfo{pages}{4763} (\bibinfo{year}{2007}).

\bibitem[{\citenamefont{McMillan et~al.}(2006)\citenamefont{McMillan, Yang,
  Panoiu, Osgood, and Wong}}]{McMillan}
\bibinfo{author}{\bibfnamefont{J.~F.} \bibnamefont{McMillan}},
  \bibinfo{author}{\bibfnamefont{X.}~\bibnamefont{Yang}},
  \bibinfo{author}{\bibfnamefont{N.~C.} \bibnamefont{Panoiu}},
  \bibinfo{author}{\bibfnamefont{R.~M.} \bibnamefont{Osgood}},
  \bibnamefont{and} \bibinfo{author}{\bibfnamefont{C.~W.} \bibnamefont{Wong}},
  \bibinfo{journal}{Opt. Lett.} \textbf{\bibinfo{volume}{31}},
  \bibinfo{pages}{1235} (\bibinfo{year}{2006}).

\bibitem[{\citenamefont{Ivchenko et~al.}(1996)\citenamefont{Ivchenko,
  Kaliteevski, Kavokin, Nesvizhskii, and Ioffe}}]{Ivchenko}
\bibinfo{author}{\bibfnamefont{E.~L.} \bibnamefont{Ivchenko}},
  \bibinfo{author}{\bibfnamefont{M.~A.} \bibnamefont{Kaliteevski}},
  \bibinfo{author}{\bibfnamefont{A.~V.} \bibnamefont{Kavokin}},
  \bibinfo{author}{\bibfnamefont{A.~I.} \bibnamefont{Nesvizhskii}},
  \bibnamefont{and} \bibinfo{author}{\bibfnamefont{A.~F.} \bibnamefont{Ioffe}},
  \bibinfo{journal}{J. Opt. Soc. Am. B} \textbf{\bibinfo{volume}{13}},
  \bibinfo{pages}{1061} (\bibinfo{year}{1996}).

\bibitem[{\citenamefont{Khitrova et~al.}(1999)\citenamefont{Khitrova, Gibbs,
  Jahnke, Kira, and Koch}}]{Khitrova}
\bibinfo{author}{\bibfnamefont{G.}~\bibnamefont{Khitrova}},
  \bibinfo{author}{\bibfnamefont{H.~M.} \bibnamefont{Gibbs}},
  \bibinfo{author}{\bibfnamefont{F.}~\bibnamefont{Jahnke}},
  \bibinfo{author}{\bibfnamefont{M.}~\bibnamefont{Kira}}, \bibnamefont{and}
  \bibinfo{author}{\bibfnamefont{S.~W.} \bibnamefont{Koch}},
  \bibinfo{journal}{Rev. Mod. Phys.} \textbf{\bibinfo{volume}{71}},
  \bibinfo{pages}{1591} (\bibinfo{year}{1999}).

\bibitem[{\citenamefont{John and Florescu}(2001)}]{John}
\bibinfo{author}{\bibfnamefont{S.}~\bibnamefont{John}} \bibnamefont{and}
  \bibinfo{author}{\bibfnamefont{M.}~\bibnamefont{Florescu}},
  \bibinfo{journal}{Journal of Optics A: Pure and Applied Optics}
  \textbf{\bibinfo{volume}{3}}, \bibinfo{pages}{S103} (\bibinfo{year}{2001}).

\bibitem[{\citenamefont{Soljacic and
  Joannopoulos}(2004)}]{Soljacic.Nat}
\bibinfo{author}{\bibfnamefont{M.}~\bibnamefont{Soljacic}} \bibnamefont{and}
  \bibinfo{author}{\bibfnamefont{J.~D.} \bibnamefont{Joannopoulos}},
  \bibinfo{journal}{Nat Mater} \textbf{\bibinfo{volume}{3}},
  \bibinfo{pages}{211} (\bibinfo{year}{2004}).

\bibitem[{\citenamefont{Fleischhauer et~al.}(2005)\citenamefont{Fleischhauer,
  Imamoglu, and Marangos}}]{Fleisch}
\bibinfo{author}{\bibfnamefont{M.}~\bibnamefont{Fleischhauer}},
  \bibinfo{author}{\bibfnamefont{A.}~\bibnamefont{Imamoglu}}, \bibnamefont{and}
  \bibinfo{author}{\bibfnamefont{J.~P.} \bibnamefont{Marangos}},
  \bibinfo{journal}{Rev. Mod. Phys.} \textbf{\bibinfo{volume}{77}},
  \bibinfo{pages}{633} (\bibinfo{year}{2005}).

\bibitem[{\citenamefont{Soljacic
  et~al.}(2005)\citenamefont{Soljacic, Lidorikis, Hau, and
  Joannopoulos}}]{Soljacic}
\bibinfo{author}{\bibfnamefont{M.}~\bibnamefont{Soljacic}},
  \bibinfo{author}{\bibfnamefont{E.}~\bibnamefont{Lidorikis}},
  \bibinfo{author}{\bibfnamefont{L.~V.} \bibnamefont{Hau}}, \bibnamefont{and}
  \bibinfo{author}{\bibfnamefont{J.~D.} \bibnamefont{Joannopoulos}},
  \bibinfo{journal}{Phys. Rev. E} \textbf{\bibinfo{volume}{71}},
  \bibinfo{pages}{026602} (\bibinfo{year}{2005}).

\bibitem[{\citenamefont{Arkhipkin and Myslivets}(2009)}]{Arkhip}
\bibinfo{author}{\bibfnamefont{V.~G.} \bibnamefont{Arkhipkin}}
  \bibnamefont{and} \bibinfo{author}{\bibfnamefont{S.~A.}
  \bibnamefont{Myslivets}}, \bibinfo{journal}{Quantum Electron}
  \textbf{\bibinfo{volume}{39}}, \bibinfo{pages}{157} (\bibinfo{year}{2009}).

\bibitem[{\citenamefont{Payne and Deng}(2001)}]{Deng}
\bibinfo{author}{\bibfnamefont{M.~G.} \bibnamefont{Payne}} \bibnamefont{and}
  \bibinfo{author}{\bibfnamefont{L.}~\bibnamefont{Deng}},
  \bibinfo{journal}{Phys. Rev. A} \textbf{\bibinfo{volume}{64}},
  \bibinfo{pages}{031802(R)} (\bibinfo{year}{2001}).

\bibitem[{\citenamefont{Inouye et~al.}(2000)\citenamefont{Inouye, L\"ow, Gupta,
  Pfau, G\"orlitz, Gustavson, Pritchard, and Ketterle}}]{Inouye}
\bibinfo{author}{\bibfnamefont{S.}~\bibnamefont{Inouye}},
  \bibinfo{author}{\bibfnamefont{R.~F.} \bibnamefont{L\"ow}},
  \bibinfo{author}{\bibfnamefont{S.}~\bibnamefont{Gupta}},
  \bibinfo{author}{\bibfnamefont{T.}~\bibnamefont{Pfau}},
  \bibinfo{author}{\bibfnamefont{A.}~\bibnamefont{G\"orlitz}},
  \bibinfo{author}{\bibfnamefont{T.~L.} \bibnamefont{Gustavson}},
  \bibinfo{author}{\bibfnamefont{D.~E.} \bibnamefont{Pritchard}},
  \bibnamefont{and} \bibinfo{author}{\bibfnamefont{W.}~\bibnamefont{Ketterle}},
  \bibinfo{journal}{Phys. Rev. Lett.} \textbf{\bibinfo{volume}{85}},
  \bibinfo{pages}{4225} (\bibinfo{year}{2000}).

\bibitem[{\citenamefont{Lee and Lawandy}(2001)}]{Lee}
\bibinfo{author}{\bibfnamefont{K.}~\bibnamefont{Lee}} \bibnamefont{and}
  \bibinfo{author}{\bibfnamefont{N.~M.} \bibnamefont{Lawandy}},
  \bibinfo{journal}{Applied Physics Letters} \textbf{\bibinfo{volume}{78}},
  \bibinfo{pages}{703} (\bibinfo{year}{2001}).

\bibitem[{\citenamefont{Sharping et~al.}(2005)\citenamefont{Sharping, Okawachi,
  and Gaeta}}]{Sharping}
\bibinfo{author}{\bibfnamefont{J.}~\bibnamefont{Sharping}},
  \bibinfo{author}{\bibfnamefont{Y.}~\bibnamefont{Okawachi}}, \bibnamefont{and}
  \bibinfo{author}{\bibfnamefont{A.}~\bibnamefont{Gaeta}},
  \bibinfo{journal}{Opt. Express} \textbf{\bibinfo{volume}{13}},
  \bibinfo{pages}{6092} (\bibinfo{year}{2005}).

\bibitem[{\citenamefont{Klyshko}(1986)}]{Klyshko}
\bibinfo{author}{\bibfnamefont{D.}~\bibnamefont{Klyshko}},
  \emph{\bibinfo{title}{Physical basics of quantum electronics [in Russian]}}
  (\bibinfo{publisher}{Moscow: Nauka}, \bibinfo{year}{1986}).

\bibitem[{\citenamefont{Akhmanov and Koroteev}(1981)}]{Akhmanov}
\bibinfo{author}{\bibfnamefont{S.~A.} \bibnamefont{Akhmanov}} \bibnamefont{and}
  \bibinfo{author}{\bibfnamefont{N.~I.} \bibnamefont{Koroteev}},
  \emph{\bibinfo{title}{Methods of Nonlinear Optics in Light Scattering
  Spectroscopy [in Russian]}} (\bibinfo{publisher}{Moscow: Nauka},
  \bibinfo{year}{1981}).

\bibitem[{\citenamefont{Repasky et~al.}(1999)\citenamefont{Repasky, Meng,
  Brasseur, Carlsten, and Swanson}}]{Repasky}
\bibinfo{author}{\bibfnamefont{K.~S.} \bibnamefont{Repasky}},
  \bibinfo{author}{\bibfnamefont{L.}~\bibnamefont{Meng}},
  \bibinfo{author}{\bibfnamefont{J.~K.} \bibnamefont{Brasseur}},
  \bibinfo{author}{\bibfnamefont{J.~L.} \bibnamefont{Carlsten}},
  \bibnamefont{and} \bibinfo{author}{\bibfnamefont{R.~C.}
  \bibnamefont{Swanson}}, \bibinfo{journal}{J. Opt. Soc. Am. B}
  \textbf{\bibinfo{volume}{16}}, \bibinfo{pages}{717} (\bibinfo{year}{1999}).

\bibitem[{\citenamefont{Poelker and Kumar}(1992)}]{Kumar}
\bibinfo{author}{\bibfnamefont{M.}~\bibnamefont{Poelker}} \bibnamefont{and}
  \bibinfo{author}{\bibfnamefont{P.}~\bibnamefont{Kumar}},
  \bibinfo{journal}{Opt. Lett.} \textbf{\bibinfo{volume}{17}},
  \bibinfo{pages}{399} (\bibinfo{year}{1992}).

\bibitem[{\citenamefont{Yariv}(1975)}]{Yariv}
\bibinfo{author}{\bibfnamefont{A.}~\bibnamefont{Yariv}},
  \emph{\bibinfo{title}{Quantum Electronics, 2d ed.}} (\bibinfo{publisher}{New
  York: Wiley}, \bibinfo{year}{1975}).

\bibitem[{\citenamefont{Boyd}(1992)}]{Boyd}
\bibinfo{author}{\bibfnamefont{R.}~\bibnamefont{Boyd}},
  \emph{\bibinfo{title}{Nonlinear optics}} (\bibinfo{publisher}{London:
  Academic Press}, \bibinfo{year}{1992}).

\bibitem[{\citenamefont{Balakin et~al.}(2001)\citenamefont{Balakin, Bushuev,
  Mantsyzov, Ozheredov, Petrov, Shkurinov, Masselin, and Mouret}}]{Balakin}
\bibinfo{author}{\bibfnamefont{A.~V.} \bibnamefont{Balakin}},
  \bibinfo{author}{\bibfnamefont{V.~A.} \bibnamefont{Bushuev}},
  \bibinfo{author}{\bibfnamefont{B.~I.} \bibnamefont{Mantsyzov}},
  \bibinfo{author}{\bibfnamefont{I.~A.} \bibnamefont{Ozheredov}},
  \bibinfo{author}{\bibfnamefont{E.~V.} \bibnamefont{Petrov}},
  \bibinfo{author}{\bibfnamefont{A.~P.} \bibnamefont{Shkurinov}},
  \bibinfo{author}{\bibfnamefont{P.}~\bibnamefont{Masselin}}, \bibnamefont{and}
  \bibinfo{author}{\bibfnamefont{G.}~\bibnamefont{Mouret}},
  \bibinfo{journal}{Phys. Rev. E} \textbf{\bibinfo{volume}{63}},
  \bibinfo{pages}{046609} (\bibinfo{year}{2001}).

\bibitem[{\citenamefont{Yariv}(1976)}]{Yariv1}
\bibinfo{author}{\bibfnamefont{A.}~\bibnamefont{Yariv}},
  \emph{\bibinfo{title}{Introduction to optical electronics, 2d ed.}}
  (\bibinfo{publisher}{New York}, \bibinfo{year}{1976}).

\bibitem[{\citenamefont{Repasky et~al.}(1998)\citenamefont{Repasky, Brasseur,
  Meng, and Carlsten}}]{Repasky1}
\bibinfo{author}{\bibfnamefont{K.~S.} \bibnamefont{Repasky}},
  \bibinfo{author}{\bibfnamefont{J.~K.} \bibnamefont{Brasseur}},
  \bibinfo{author}{\bibfnamefont{L.}~\bibnamefont{Meng}}, \bibnamefont{and}
  \bibinfo{author}{\bibfnamefont{J.~L.} \bibnamefont{Carlsten}},
  \bibinfo{journal}{J. Opt. Soc. Am. B} \textbf{\bibinfo{volume}{15}},
  \bibinfo{pages}{1667} (\bibinfo{year}{1998}).

\bibitem[{\citenamefont{Shimizu et~al.}(2002)\citenamefont{Shimizu, Shiokawa,
  Yamamoto, Kozuma, Kuga, Deng, and Hagley}}]{Shimizu}
\bibinfo{author}{\bibfnamefont{Y.}~\bibnamefont{Shimizu}},
  \bibinfo{author}{\bibfnamefont{N.}~\bibnamefont{Shiokawa}},
  \bibinfo{author}{\bibfnamefont{N.}~\bibnamefont{Yamamoto}},
  \bibinfo{author}{\bibfnamefont{M.}~\bibnamefont{Kozuma}},
  \bibinfo{author}{\bibfnamefont{T.}~\bibnamefont{Kuga}},
  \bibinfo{author}{\bibfnamefont{L.}~\bibnamefont{Deng}}, \bibnamefont{and}
  \bibinfo{author}{\bibfnamefont{E.~W.} \bibnamefont{Hagley}},
  \bibinfo{journal}{Phys. Rev. Lett.} \textbf{\bibinfo{volume}{89}},
  \bibinfo{pages}{233001} (\bibinfo{year}{2002}).

\bibitem[{\citenamefont{Zhang et~al.}(2008)\citenamefont{Zhang, Hernandez, and
  Zhu}}]{Zhang}
\bibinfo{author}{\bibfnamefont{J.}~\bibnamefont{Zhang}},
  \bibinfo{author}{\bibfnamefont{G.}~\bibnamefont{Hernandez}},
  \bibnamefont{and} \bibinfo{author}{\bibfnamefont{Y.}~\bibnamefont{Zhu}},
  \bibinfo{journal}{Opt. Lett.} \textbf{\bibinfo{volume}{33}},
  \bibinfo{pages}{46} (\bibinfo{year}{2008}).

\end{thebibliography}

\end{document}